\documentclass[a4paper,11pt]{article}
\usepackage{graphicx}
\usepackage{amsmath,amssymb,bm,dcolumn}

\newcommand{\be}{\begin{equation}}
\newcommand{\ee}{\end{equation}}
\newcommand{\bea}{\begin{eqnarray}}
\newcommand{\eea}{\end{eqnarray}}

\begin{document}

\thispagestyle{empty}
\begin{center}
\Large{\bf Quantum Field Theory Approach to the Optical Conductivity of Strained and Deformed Graphene.} \\
\vspace{1cm}
\Large{W. de Paula$^{1}$, A. Chaves$^{2}$, O. Oliveira$^{3}$ and T. Frederico$^{1}$}

\center{
\small{\small{1-Departamento de F\'isica, Instituto Tecnol\'ogico de Aeron\'autica}},\\ 
\small{\small{2- Department of Physics and Center of Physics, University of Minho, P-4710-057, Braga, Portugal}},\\ 
\small{\small{3- CFisUC, Department of Physics, University of Coimbra, P - 3004 516 Coimbra, Portugal}}
}

\end{center}






\begin{abstract}
The computation of the optical conductivity of strained and deformed graphene is discussed within the framework of quantum field
theory in curved spaces. The analytical solutions of the Dirac equation in an arbitrary static background geometry for one dimensional
periodic deformations are computed, together with the corresponding Dirac propagator. Analytical expressions are given for the optical
conductivity of strained and deformed graphene associated with both intra and interbrand transitions. The special case of small
deformations is discussed and the result compared to the prediction of the tight-binding model.

\end{abstract}

\section{Introduction}

Graphene is a 2D material whose low energy electronic excitations can be described by a massless
Dirac equation~\cite{Neto2009,Peres2010,Chaves:2010zn,Oliveira:2010hq,Popovici:2012xs,Chaves:2013fca}. The Dirac equation can be motivated starting from the tight-binding model and assuming
that graphene is a flat sheet.  However, almost immediately after its discovery deviations from  flatland were observed with the identification
of ripples and corrugations. Further, in many experimental realisations, graphene is grown on top of a supporting material which is hardly flat.
On the theoretical side, the analysis of the stability of two-dimensional crystals and membranes suggests that flat materials are not realised in nature.

The introduction of a Dirac equation to describe graphene opened the possibility of using Relativistic Field Theory to investigate its
properties. In this sense, graphene provides a nice laboratory where phenomena typically associated with High Energy Physics can be
observed and studied. Deviations from the perfect cristal structure can easily be accommodated within the field theoretical approach
calling for new fields and/or operators. Indeed, it was proven that certain types of defects require the introduction of a scalar field and/or
two dimensional gauge fields. If these fields are dynamical, phenomena like spontaneous symmetry breaking are
expected to occur in graphene. Moreover, these relativistic field theories have its proper dynamics, with e.g. soliton like solutions \cite{Cordeiro:2013vla},
which couple to the electrons and can induce non-trivial effects. The full picture of the coupling between the various possible fields
is still to be explored.

On the other side, geometric deformations of graphene are known to change its electrical conductance and optical properties.
From the point of view of the electronic properties and the field theoretical approach the natural language to describe
deformed graphene is via the Dirac equation in a curved space. Once more, graphene makes a new bridge between two
distinct and far apart branches of physics. Recall that, electrons are confined within a 2D surface which leaves in a 3D world.
Further, as discussed in \cite{Chaves:2013fca} the use of a Dirac equation in a 2D curved space can accommodate not only geometric
deformations of the pristine material but can also be used to study strained graphene.

As described in \cite{Vozmediano2010}, certain types of geometric deformation have associated intense pseudo magnetic fields. The interpretation of Landau levels spacing in terms of a pseudo magnetic field unveil intense fields of the order $\sim$300 T.
For one dimensional periodic deformations like e.g. ripples, in~\cite{Chaves:2013fca} it was studied how optical properties are changed by
deforming graphene.
The main goal of all such studies is, of course, to learn how to tailor the properties of such a promising material as graphene.

This paper is organised as follows. In section \ref{curved_space} we resume the main features of the Dirac equation in a static background geometry and discuss how
strain and geometric deformations can be described within such a formalism. In section \ref{Sec:1Ddef}, we discuss in detail the case of one dimensional deformations
and write out the expressions for the electronic Green function. In section \ref{opt_cond}, the report on the computation of the optical conductivity for the Dirac equation
in curved space using the Kubo formula.

\section{Dirac Electron in Curved Space \label{curved_space}}

The Dirac equation in a curved space-time with a metric tensor $g_{\mu\nu}$ reads
\be
  \Big\{ i \, \Gamma^{\mu} D_{\mu} - M \Big\}\Psi=0
  \label{Eq:Dirac}
\ee
where $\Psi (x)$ is the Dirac field, $\Gamma^{\mu}=e^{\mu}_{A}\gamma^{A}$ are the Dirac matrices in curved space-time, where the
 ``vielben" $e^{\mu}_{A}$ defines a local Lorentzian frame such that the space-time interval is given by
$ds^2 = \, \eta_{AB} \,  \theta^{A} \, \theta^{B}$;  $\eta_{AB} = diag(-1, 1, 1)$ is the Minkowski metric and $\theta^{A} = e^A_\mu \, dx^\mu$.
The covariant derivative is
\be D_{\mu} = \partial_{\mu} + \frac{1}{4}\omega^{AB}_{\mu}\sigma_{AB} \qquad\mbox{ with }\qquad
 \sigma_{AB} = \frac{1}{2} \Big(\gamma_{A}\gamma_{B}-\gamma_{B}\gamma_{A}\Big)
\ee
and the spin connection are given by
\be
\omega_{\mu}^{AB} = \frac{1}{2} \, e^{\nu A} \, \Big(\partial_{\mu}e^{B}_{\nu} \, - \, \partial_{\nu}e^{B}_{\mu}\Big)
       \, - \, \frac{1}{2}e^{\nu B}\Big(\partial_{\mu}e^{A}_{\nu} \, - \, \partial_{\nu}e^{A}_{\mu}\Big)
       \, - \,  \frac{1}{2}e^{\rho A}e^{\sigma B}\Big(\partial_{\rho}e_{\sigma C}
      \, - \, \partial_{\sigma}e_{\rho C} \Big) e^{C}_{\mu}.
\ee

For a time independent metric one can set $\theta^0  = dt$ and choose $\Gamma^0 = \gamma^0$, where $\gamma^0$ is the usual flatland gamma matrix.
Furthermore, from the definition
$\theta^1  =   e^1_1 \, dx ~ + ~ e^1_2 \, dy \ , \theta^2  =   e^2_1 \, dx ~ + ~ e^2_2 \, dy$ and taking the symmetric solution $e^1_2  = e^2_1$, it follows, after some algebra,
that and the only non-vanishing spin connections are $\omega^{12}_i = - \omega^{21}_i$ for $i = 1$, $2$. The Dirac equation (\ref{Eq:Dirac}) can be written in the usual form
\be
   i \partial_t \Psi = H \, \Psi
\ee
where the Hamiltonian is
\be
  H =  - i \sum^2_{i,j = 1} \alpha_i \, e^j_i \, \nabla_j ~ -  ~     i \sum^2_{i=1} \alpha_i A_i + M \beta \ ,
    \label{Eq:H}
\ee
with
\be
 \beta = \gamma^0 \ , \qquad \alpha_i = \beta \gamma^i \ ,  \qquad A_i  =     \sum^2_{j,k=1} \, \frac{\epsilon_{ij}}{2} \, e^k_j \, \omega^{12}_k
  \label{Eq:pseudo_gauge}
\ee
and $\epsilon_{ij}$ is the two dimensional Levi-Civita symbol.
Equation (\ref{Eq:H}) extends, for a generic static geometry, the Hamiltonian derived in \cite{Juan2007,Juan2012} for radial and small deformations;
see also~\cite{Kerner2012}. Following the notation of \cite{Juan2012}, the first term in $H$ includes the usual flat Dirac term and the space-dependent Fermi velocity
given by the departure from unit of the corresponding "vielben". The second term can be interpreted as a geometric gauge field. Note, however,
the absence of the usual $i$ factor appearing in the minimal U(1) coupling.
The last term, is a mass term and it shows up only when chiral symmetry is explicitly broken and, therefore, one can conclude that in graphene a gap is not open
by a geometric deformation.
%
%

We define the geometric deformation of the graphene sheet according to the parametric equation
\begin{equation}
   z = h(x,y) \ .
   \label{Eq:outplane}
\end{equation}
Strain can be viewed as a local deformation and can be parametrized by the change of variables
\begin{equation}
  x \rightarrow x^\prime = \mathcal{X}(x,y)  =  x + X_x(x,y) \qquad\mbox{and}\qquad y \rightarrow y^\prime = \mathcal{Y}(x,y)  = y + Y_y(x,y)\ ,
  \label{Eq:strain}
\end{equation}
where in terms of $(x^\prime, y^\prime)$ the metric is euclidean and $X_x$, $Y_y$ and $h$ are the displacement fields.
The transformation (\ref{Eq:strain}) does not change the surface curvature,
in the sense that the 2D Ricci scalar defined on the graphene sheet is invariant under the change of variables. The distance between two neighboring points
is given by
\begin{equation}
  ds^2 = \left( dx^\prime \right)^2 + \left( dy^\prime \right)^2  + dz^2
\end{equation}
and the corresponding metric tensor reads $g_{ij}  = \mathcal{X}_i \mathcal{X}_j \ + \ \mathcal{Y}_i \mathcal{Y}_j \ + \  h_i h_j$,
where $\mathcal{X}_i$ means partial derivative with respect to $x_i$, and can be written as
\begin{equation}
  g_{ij}   =  \delta_{ij}  + 2 \, u_{ij}
  \label{Eq:metric}
\end{equation}
where
\be
u_{ij} = \frac{1}{2}\left( \partial_i u_j + \partial_j u_i + \sum^2_{k=1} \partial_i u_k  \  \partial_j u_k
    + \left( \partial_i h \right) \left( \partial_j h \right) \right) \
    \label{Eq:straintensor}
\ee
is the strain tensor and $u_x = X_x$, $u_y = Y_y$.
In the limit of the small deformations and to first order in the in-plane displacements and to second
order in the out-of-plane displacement, the strain tensor  $u_{ij}$ reproduces exactly the strain
tensor considered in \cite{Juan2012}. The difference for a finite deformation is the second order term
in the in-plane displacements $X_x$ and $Y_y$.

It follows from equations (\ref{Eq:metric}) and (\ref{Eq:straintensor}) that strain and geometric deformations play similares roles. Indeed, as
long as the stress tensor is the same, there is no basic difference between strain and deformation. Furthermore, for a particular geometric
deformation there is, in principle, a set of strains such that $u_{ij} = 0$ and the combined effect is flat graphene. In the same way, strained
graphene can behave as flat graphene provided it is deformed conveniently production a vanishing stress tensor. One should not forget
that under sufficiently large deformations, the overlap between the various electron orbitals can change significantly and change the nature
of the electronic interaction and new interactions are required in the Dirac equation. In this sense, a geometric deformation is not equivalent
to straining graphene.


\section{One Dimensional Deformations \label{Sec:1Ddef}}

Let us consider  the case of one dimensional deformations defined by $z=h(x)$ and $X_x(x,y)=X_x(x)$.
A deformation on the $x$ direction induces a deformation $y$. The stretch along $y$ is related to the stretch in $x$ direction via
the Poisson ratio and in this sense one can expect that $Y_y(x,y)=Y_y(x) y$. However, in order to simplify the calculations
we will take $Y_y (x,y) = Y_y(x)$. The corresponding Dirac Hamiltonian reads \cite{Chaves:2013fca}:
\be
{\cal H}=-i\frac{\alpha_x}{\sqrt{g(x)}}\partial_x -2i\frac{u_{xy}(x)}{\sqrt{g(x)}} \partial_y-i\alpha_y\partial_y,\label{eq:1D}
\ee
where
\be
\sqrt{g(x)}=\sqrt{1+2u_{xx}-4u_{xy}^2}
\ee
and whose eigenvectors are given by
\be
\Psi_{k_x,k_y,\lambda}(x,y)=\frac{1}{\sqrt{2\xi L}} e^{ik_y(y+Y_y(x))}e^{ik_x\int^x dx^\prime \sqrt{g(x^\prime)}} \begin{pmatrix}1\\ i\lambda e^{-i\theta} \end{pmatrix},
\label{Eq:SolPsi}
\ee
where $\theta=\text{atan} (k_y/k_x)$, $\lambda=\pm 1$. The energy of solution (\ref{Eq:SolPsi}) is $E=\lambda\sqrt{k_x^2+k_y^2}$.


Electrons in the conduction band are described by the solutions (\ref{Eq:SolPsi}) with $\lambda = +1$, while those electrons in the valence band
are described by $\lambda = -1$. In the following, we will assume that the valence band is occupied. The Dirac propagator in coordinate
space reads
%
%
\be
G(x,y,x^\prime,y^\prime,E)=\sum_k \frac{\Psi_{k_x,k_y,+}(x,y)\Psi^\dagger_{k_x,k_y,+}(x^\prime,y^\prime)}{E-E_k+i\varepsilon}+\frac{\Psi_{k_x,k_y,-}(x,y)\Psi^\dagger_{k_x,k_y,-}(x^\prime,y^\prime)}{E+E_k-i\varepsilon},
\label{Eq:DiracProp}
\ee
where $E_k=\sqrt{k_x^2+k_y^2}$. Assuming periodic boundary conditions in $x$ and $y$ direction, i.e.
%
%
\begin{flalign}
X_x(-L/2) = X_x(L/2) \qquad \mbox{ and } \qquad Y_y(-L/2)=Y_y(L/2) ,
\end{flalign}
then
\begin{flalign}
\sum_{k_y} \rightarrow \frac{L}{2\pi} \int dk_y  \qquad\mbox{ and }\qquad \sum_{k_x}\rightarrow \frac{\xi}{2\pi}\int dk_x
\end{flalign}
where we have defined
\be
\xi=\int_{-L/2}^{L/2} d x\sqrt{g(x)} \ ,
\ee
then the Green function (\ref{Eq:DiracProp}) can be rewritten as
\bea
G(x,y,x^\prime,y^\prime,E) & = &
   \int dk_x	 dk_y \frac{e^{ik_y(y+Y_y(x)-y^\prime-Y_y(x^\prime))}e^{ik_x\int_{x^\prime}^x dx^{\prime\prime}\sqrt{g(x^{\prime\prime})}}}{8\pi^2} \nonumber \\
   & & \qquad\qquad\qquad \left[\frac{M_+^\theta}{E-E_k+i\varepsilon}+\frac{M_-^\theta}{E+E_k-i\varepsilon}\right]
\eea
with
\be
M^\theta_\lambda=\begin{pmatrix} 1 && -i\lambda e^{i\theta} \\ i\lambda e^{-i\theta} &&1\end{pmatrix}.
\ee
In terms of $x$, $x^\prime$ and the momentum along direction $y$, the later expression becomes
\be
G(x,x^\prime,k_y,E)=\frac{e^{ik_y(Y_y(x)-Y_y(x^\prime))}}{8\pi^2}\int dk_x  e^{ik_x\int_{x^\prime}^x dx^{\prime\prime}\sqrt{g(x^{\prime\prime})}}\left[\frac{M_+^\theta}{E-E_k+i\varepsilon}+\frac{M_-^\theta}{E+E_k-i\varepsilon}\right].
\ee
The integration over $k_x$ can be performed using standard techniques and gives
\be
G(x,x^\prime,k_y,E)=\frac{e^{ik_y(Y_y(x)-Y_y(x^\prime))}}{4\pi i} \, \frac{E}{k_0}\, e^{ik_0|\zeta(x,x^\prime)|}
\, \begin{pmatrix}1 && -i e^{i\,\text{sgn}\left[\zeta(x,x^\prime)\right]\theta}\\ i e^{-i\,\text{sgn}\left[\zeta(x,x^\prime)\right]\theta}&& 1 \end{pmatrix},
\ee
where
\be
\zeta(x,x^\prime)=\int_{x^\prime}^x dx^{\prime\prime}\sqrt{g(x^{\prime\prime})},
\qquad\mbox{ and }\qquad
k_0=\sqrt{E^2-k_y^2}.
\ee
Note that all information on the geometric of the graphene sheet is resumed in the function $\zeta (x, x^\prime)$. Strain shows up in $\zeta$
and in the extra phase $\delta_y = k_y(Y_y(x)-Y_y(x^\prime))$ which vanishes if the graphene is dislocated uniformly along $y$.

\section{The Optical Conductivity \label{opt_cond}}

The active electrons contributing to the optical conductivity can belong to the valence band, which are described by the negative energy eigenstates of
$H$, and to the conduction band, which are the positive energy eigenstates of the hamiltonian. One can introduce a chemical potential
$\mu$ associated to the occupation of the electronic bands. A negative chemical potential means a partially occupied valence band.
On the other hand, a $\mu>0$  describes a partially filled conduction band.

In our particle independent model, the optical conductivity of deformed and strained graphene can be computed via the Kubo
formula~\cite{Mahan,HipPRB12} adapted to a two dimension problem
\be
\sigma_{ij} = - i \, \mathcal{G} \, \sum_{m,n}\frac{f(E_m-\mu) - f(E_n-\mu)}{\omega-\omega_{mn}+i\epsilon} \, v^i_{mn} \, v^j_{nm}, \label{kubo}
\ee
where $\mathcal{G} = \frac{e^2g_s}{A \omega}$ , $e$ is the electron electric charge, $g_s = 4$ is the spin and pseudospin degenerescence,
$A=\xi L$ is the effective area of the graphene sheet, $\Omega$ is the light frequency, $\Omega_{mn} = E_m - E_n$ is the transition energy,
$f(\varepsilon)=(1+e^{\beta \, \varepsilon})^{-1}$ is the Fermi-Dirac distribution function, $\beta=1/k_BT$ and
$v^i_{mn}=\langle m|v_i| n\rangle$ are the matrix elements of the velocity operator in the Hamiltonian eigenvector basis.

The  velocity operators are defined as $v_i = i \, \left[ H , x_i \right]= i \, \alpha_j \, e^j_i$, where $H$ is the hamiltonian reported in (\ref{Eq:H}).
From the definition it follows that
\be
v_x \, = \, \frac{\alpha_x}{\sqrt{g}} \qquad\mbox{ and } \qquad
v_y \, = \, \alpha_y-\frac{2\, u_{xy}}{\sqrt{g}}\, \alpha_x
             \label{Eq:vel}
\ee
with $\vec\alpha = \beta \, \vec\gamma$.

The real part of the longitudinal optical conductivity $\sigma_{ii}$ is associated with absorption and it reads
\be
\Re \, \sigma _{ii}  = - \mathcal{G} \, \pi \, \sum_{m,n}\Big[f(E_m-\mu)-f(E_n-\mu)\Big]\delta\left(\omega-\omega_{mn}\right)\,|v^i_{mn}| ^2 \label{kubo1}
\ee
where the sum over states is constrained by energy conservation.

The computation of the longitudinal optical conductivity require the matrix elements of the velocity operator
\begin{flalign}
&
\langle k_x^\prime,k_y^\prime,\lambda^\prime |v_x | k_x,k_y,\lambda\rangle  = -\delta_{k_y,k_y^\prime} F(\Delta k_x)\frac{\lambda e^{-i\theta}+\lambda^\prime e^{i\theta^\prime}}{2},
\\
&
\langle k_x^\prime,k_y^\prime,\lambda^\prime |v_y | k_x,k_y,\lambda\rangle  =  i\delta_{k_y,k_y^\prime}\delta_{k_x,k_x^\prime} \frac{\lambda e^{-i\theta}-\lambda^\prime e^{i\theta^\prime} }{2}
+2\delta_{k_y,k_y^\prime}G(\Delta k_x)\frac{\lambda e^{-i\theta}+\lambda^\prime e^{i\theta^\prime} }{2},
\end{flalign}
where
\begin{flalign}
& F(p_x)=\frac{1}{\xi} \int_{-L/2}^{L/2} d x ~ \exp\left[i \, p_x \, \int^x d x^\prime \sqrt{g(x^\prime)}\right],
\\
& G(p_x)=\frac{1}{\xi} \int_{-L/2}^{L/2} d x ~ u_{xy} ~ \exp\left[i \, p_x \, \int^x d x^\prime \sqrt{g(x^\prime)}\right],
\end{flalign}
The electronic transitions where matrix elements have the same $\lambda$ in the initial and final state are called intraband transitions.
On the other hand, those transitions where $\lambda$ changes sign are called interband transitions.


Introducing the following notation for the difference of Fermi-Dirac functions
\be
\Delta_{k_x,k_x^\prime,k_y}^{\lambda\lambda^\prime}(\mu)=n_F\left(\lambda \sqrt{k_x^2+k_y^2}-\mu\right)-n_F\left(\lambda^\prime \sqrt{{k_x^\prime}^2+k_y^2}-\mu\right) \ ,
\ee
then the conductivity is given by
\be
\frac{{\cal R}\sigma_{xx}^\text{inter}}{\sigma_0}=-\frac{2\, \xi}{\pi^2\omega}\int dk_x \int dk_x^\prime \int dk_y ~ \Delta_{k_x,k_x^\prime,k_y}^{+-}(\mu) ~
\delta(\omega-\omega_{mn}) ~  |F(\Delta k_x)|^2  ~  \sin^2\left(\frac{\theta+\theta^\prime}{2}\right),
\ee
\be
\frac{{\cal R}\sigma_{xx}^\text{intra}}{\sigma_0}= -\frac{2\, \xi}{\pi^2\omega}\sum_\lambda\int dk_x \int dk_x^\prime \int dk_y  ~ \Delta_{k_x,k_x^\prime,k_y}^{\lambda\lambda}(\mu)~ \delta(\omega-\omega_{mn}) ~ |F(\Delta k_x)|^2  ~ \cos^2\left(\frac{\theta+\theta^\prime}{2}\right),
\ee
\bea
\frac{{\cal R}\sigma_{yy}^\text{inter}}{\sigma_0} & = & -\frac{2\,\xi}{\pi^2\omega}\int dk_x \int dk_x^\prime \int dk_y ~ \Delta_{k_x,k_x^\prime,k_y}^{+-}(\mu)  ~
\delta(\omega-\omega_{mn}) ~ \Bigg[ \delta_{k_x,k_x^\prime} \cos^2\left(\frac{\theta+\theta^\prime}{2}\right) +\nonumber\\
& & \qquad\qquad\qquad
+ ~ 4 \, |G(\Delta k_x)|^2 ~ \sin^2\left(\frac{\theta+\theta^\prime}{2}\right)  -  \delta_{k_x,k_x^\prime} ~ {\cal R}\{G(\Delta k_x)\} ~ \sin(\theta+\theta^\prime) \Bigg],\nonumber\\
\eea
\be
\frac{{\cal R}\sigma_{yy}^\text{intra}}{\sigma_0} = -\frac{2\,\xi}{\pi^2\omega}\sum_\lambda\int dk_x \int dk_x^\prime \int dk_y ~
  \Delta_{k_x,k_x^\prime,k_y}^{\lambda\lambda}(\mu)) ~ \delta(\omega-\omega_{mn})~ 4 \, |G(\Delta k_x)|^2 ~ \cos^2\left(\frac{\theta+\theta^\prime}{2}\right),
\ee
where
\be
\sigma_0=\frac{e^2}{4},
\ee
is the optical conductivity of flat graphene \cite{Peres2010}. The Dirac delta function express the conservation of energy and can be used to integrate over
$k_y$ giving
\be
\frac{{\cal R}\sigma_{xx}^\text{inter}}{\sigma_0} ~ = ~ \frac{\xi}{\pi^2\omega}\int dk_x \int dk_x^\prime  ~ \Delta_{k_x,k_x^\prime}^{+-}(\mu) ~
|F(\Delta k_x)|^2  ~ \Omega ( \omega, k_x k_x^\prime ) ,\label{eq1s} 
\ee
\be
\frac{{\cal R}\sigma_{xx}^\text{intra}}{\sigma_0} ~ = ~ -\frac{2\,\xi}{\pi^2\omega}\sum_\lambda\int dk_x \int dk_x^\prime ~ \Delta_{k_x,k_x^\prime}^{\lambda\lambda}(\mu)
~ |F(\Delta k_x)|^2  ~ \Big[1-\Omega ( \omega, k_x k_x^\prime )\Big], \label{eq2s}
\ee
\bea
\frac{{\cal R}\sigma_{yy}^\text{inter}}{\sigma_0} ~ = ~ \Delta_{\frac{\omega}{2},\frac{\omega}{2}}^{+,-}(\mu) ~ -  ~
\frac{8\,\xi}{\pi^2\omega} \int dk_x \int dk_x^\prime ~ \Delta_{k_x,k_x^\prime}^{+-}(\mu) ~  |G(\Delta k_x)|^2
~ \Omega ( \omega, k_x k_x^\prime ) , \label{eq3s}
\eea
\be
\frac{{\cal R}\sigma_{yy}^\text{intra}}{\sigma_0} ~ = ~ -\frac{8\,\xi}{\pi^2\omega} \sum_\lambda\int dk_x \int dk_x^\prime
~ \Delta_{k_x,k_x^\prime,k_y}^{\lambda\lambda}(\mu)) ~ |G(\Delta k_x)|^2 ~
\Big[1-\Omega ( \omega, k_x k_x^\prime )\Big]. \label{eq4s}
\ee
where
\be
\Omega ( \omega, k_x k_x^\prime ) =  \sqrt{\frac{\omega^2-(k_x+k_x^\prime)^2}{\omega^2-(k_x-k_x^\prime)^2}} \ .
\ee

\subsection{Special case - half-filling and zero temperature and the limit of small deformations}

For half-filling, i.e. for $\mu=0$, and at zero temperature the Fermi-Dirac are such that the above expressions can be further simplified.
The function $\Delta^{+-}$  showing in the interband case become $1$. In what concerns the intraband transitions, the $\Delta^{\lambda\lambda}$
vanish. Therefore, for this very special case only ${\cal R} \sigma_{xx}^\text{inter}$ and ${\cal R} \sigma_{yy}^\text{inter}$ survive.
In center of mass coordinates $k_m=(k_x+k_x^\prime)/2$ and $q=k_x-k_x^\prime$ and after integrationg over $k_m$ we arrive at
\bea
\frac{{\cal R}\sigma_{xx}^\text{inter}}{\sigma_0}=\frac{\xi\omega}{4\pi}\int_{-\omega}^{\omega} dq \frac{| F(q)|^2}{\sqrt{\omega^2-q^2}}, 
\hspace{2cm}
\frac{{\cal R}\sigma_{yy}^\text{inter}}{\sigma_0}=1-\frac{2\xi\omega}{\pi}\int_{-\omega}^{\omega} dq \frac{| G(q)|^2}{\sqrt{\omega^2-q^2}}.
\eea
The same expression hold if in Eqs. (\ref{eq1s})-(\ref{eq4s}) the limit $\omega\rightarrow\infty$ is taken.

For small deformations, the expressions for the optical condutivity Eqs. (\ref{eq1s})-(\ref{eq4s}) can be simplified. Considering only the first
order term in the stress tensor, the optical conductivity reads
\bea
\frac{{\cal R}\sigma_{xx}^\text{inter}}{\sigma_0}=1-\bar{u}, \label{sig1} 
\hspace{2cm}
\frac{{\cal R}\sigma_{yy}^\text{inter}}{\sigma_0}=1, \label{sig2}
\eea
where $\bar{u}$ is the mean value of the strain tensor $u_{xx}$:
\be
\bar{u}_{xx}=\lim_{L\rightarrow\infty}\frac{1}{L} \int^{L/2}_{L/2}dx \,u_{xx}(x); \label{sig3}
\ee
note that the dependence on $u_{xy}$ is of second order and so it is neglected.

Our prediction for the optical conductivity in the limit of small deformations  (\ref{sig1}) can be compared to the tight-binding model result
\cite{Pereira2010}
\be
  \frac{{\cal R}\sigma_{xx}^\text{inter}}{\sigma_0}= 1- 4 \epsilon \ , 
\hspace{2cm}
  \frac{{\cal R}\sigma_{yy}^\text{inter}}{\sigma_0}= 1 + 4 \epsilon \ , 
\ee
where $\epsilon$ is the uniaxial strain, which in the small deformation limit, determines the full strain tensor; see \cite{Pereira2010} for details.
If the tight-binding results predicts deviations from the flat graphene conductivity, (\ref{sig1}) reports a deviation from $\sigma_0$ only for
the longitudinal component ${\cal R}\sigma_{xx}^\text{inter}$. The difference in the results probably comes from ignoring the $y$ dependence
in the dislocation $Y_y(x,y)$ in calculation reported here.


\section{Summary \label{ultima_sec}}

The present reports on the use of the geometric language to investigate electrons in strained and deformed graphene and discuss the computation
the optical conductivity in this framework. The case of a general one dimensional periodic deformation is studied and the exact solutions of the Dirac equation
in an arbitrary static background geometric are computed. This allows us to compute exactly the corresponding Dirac propagator.
Furthermore, from the Kubo formula one access the optical conductivity of graphene for an arbitrary static geometry of the graphene sheet.

We call the reader attention that experimental realisation of deformations along one direction were already achieved by e.g. bending
graphene nanoribbons \cite{Koch2012}. Within the formalism of the Dirac equation in a curved background geometry one can also describe
strained graphene. Of course, the goal is to understand and control the geometry of graphene wafers in a way that hopefully, in the future,
it is possible to engineer its properties based on pure geometric deformations and/or by applying appropriate strain fields.

\section*{Acknowledgements}

The authors acknowledge financial support from the Brazilian
agencies FAPESP (Funda\c c\~ao de Amparo \`a Pesquisa do Estado de
S\~ao Paulo) and CNPq (Conselho Nacional de Desenvolvimento
Cient\'ifico e Tecnol\'ogico).



\end{document}